\documentclass[aps,prl,twocolumn,groupedaddress,showpacs]{revtex4}

\usepackage{bm}
\usepackage{graphicx}

\begin{document}

\title{Two pseudogaps with different energy scales at the antinode of the high-temperature Bi$_2$Sr$_2$CuO$_6$ superconductor using angle-resolved photoemission spectroscopy}

\author{
	K. Nakayama,$^1$
	T. Sato,$^1$
	Y.-M. Xu,$^{2,\ast}$
	Z.-H. Pan,$^{2,\dagger}$
	P. Richard,$^{3,4}$
	H. Ding,$^{2,4}$
	H.-H. Wen,$^{4,\ddagger}$
	K. Kudo,$^{5,\S}$
	T. Sasaki,$^5$
	N. Kobayashi$^5$
	and T. Takahashi$^{1,3}$
	}

\affiliation{$^1$Department of Physics, Tohoku University, Sendai 980-8578, Japan}
\affiliation{$^2$Department of Physics, Boston College, Chestnut Hill, Massachusetts 02467, USA}
\affiliation{$^3$World Premier International Research Center, Advanced Institute for Materials Research, Tohoku University, Sendai 980-8577, Japan}
\affiliation{$^4$Beijing National Laboratory for Condensed Matter Physics, and Institute of Physics, Chinese Academy of Sciences, Beijing 100190, China}
\affiliation{$^5$Institute for Materials Research, Tohoku University, Sendai 980-8577, Japan}

\date{\today}

\begin{abstract}
We have performed high-resolution angle-resolved photoemission spectroscopy on single-layered cuprate Bi$_2$Sr$_2$CuO$_6$ to clarify the origin of the pseudogap.  By using various photon energies, we have succeeded in directly observing two different pseudogaps with two different energy scales which coexist in the antinodal region: one reflects the $d_{x^2-y^2}$-wave pairing strength while the other has a larger energy scale suggesting an origin distinct from superconductivity.  The observed two-pseudogap behavior provides a key to fully understand the pseudogap phenomena in cuprates.
\end{abstract}


\maketitle

\section{I. Introduction}
The pseudogap observed in the excitation spectrum as a suppression of spectral weight in the normal state of cuprate superconductors \cite{Hong} has attracted much attention since it is closely related to the mechanism of high-$T_{\rm c}$ (transition temperature) superconductivity.  The opening of the pseudogap has been interpreted as either a precursor of Cooper pairing above $T_{\rm c}$ without phase coherence \cite{Emery} or as the development of some sort of ordered state which competes with superconductivity \cite{Varma, Chakravarty, Das}.  However, in spite of intensive studies, the origin of the pseudogap is still highly controversial.  This is largely due to the lack of consensus on the energy scale of the pseudogap.  Some experiments pointed out that the pseudogap has a different energy scale from that of the superconducting (SC) gap, indicative of the presence of two energy scales (possibly two distinct energy gaps) in the SC state \cite{Tanaka, Kondo1, Terashima, Ma, Tacon, Boyer}.  This two-gap behavior suggests that the pseudogap has a competing nature and is not directly related to superconductivity.  It has been reported that the two-gap behavior is pronounced in low-$T_{\rm c}$ systems such as heavily underdoped Bi$_2$Sr$_2$CaCu$_2$O$_8$ (Bi2212), single-layered Bi$_2$Sr$_2$CuO$_6$ (Bi2201), and La$_{2-x}$Sr$_x$CuO$_4$ (LSCO) \cite{Tanaka, Kondo1, Terashima, Ma, Tacon, Boyer}.  On the other hand, even in the low-$T_{\rm c}$ systems, there are some recent experimental studies reporting the presence of single energy scale where the SC gap below $T_{\rm c}$ and the pseudogap above $T_{\rm c}$ show an identical energy scale with no evidence for the two-gap behavior \cite{Utpal, Feng, Shi, Zhou, Nakayama}, strongly supporting a pairing origin of the pseudogap.  The apparent contradiction requires further experimental investigation on the energy scale of the pseudogap in low-$T_{\rm c}$ cuprates to elucidate the origin of the pseudogap.

In this paper, we report high-resolution angle-resolved photoemission spectroscopy (ARPES) results on single-layered cuprate Bi2201.  By comparing ARPES data obtained with two different photon energies (8.437 and 21.218 eV), we clearly found two energy scales at the antinode below $T_{\rm c}$.  We demonstrate that these energy scales persist even above $T_{\rm c}$, suggesting the presence of two different types of pseudogaps coexisting in the same momentum ($k$) region.  We discuss the implications of the present experimental results in relation to the existing models as well as the origin of the pseudogap.

\section{II. EXPERIMENTS}
High-quality single crystals of slightly overdoped (Bi,Pb)$_2$Sr$_2$CuO$_{6+\delta}$ (Pb-Bi2201; $T_{\rm c}$ $\sim$ 21 K) and nearly-optimally doped Bi$_2$Sr$_{1.6}$La$_{0.4}$CuO$_{6+\delta}$ (La-Bi2201; $T_{\rm c}$ $\sim$ 32 K) were grown by the floating-zone \cite{Kudo1, Kudo2} and the traveling-solvent floating-zone methods \cite{Wen}, respectively.  High-resolution ARPES measurements were performed using VG-SCIENTA SES2002 and MBS A1 photoemission spectrometers with xenon (Xe) and helium (He) plasma discharge lamps \cite{Souma}.  We used one of the Xe-I lines ($h\nu$ = 8.437 eV) and the He-I$\alpha$ line (21.218 eV) to excite photoelectrons.  The energy resolution was set at 2-4 meV and 6-12 meV for the measurements with the Xe and He lamps, respectively.  The angular resolution was set at 0.2$^{\circ}$.  We cleaved samples under ultrahigh vacuum better than 4$\times$10$^{-11}$ Torr to obtain a clean and fresh sample surface for ARPES measurements.  The Fermi level ($E_{\rm F}$) of samples was referenced to that of a gold film evaporated onto the sample holder.

\section{III. RESULTS AND DISCUSSION}

\begin{figure}[!t]
\begin{center}
\includegraphics[width=3.4in]{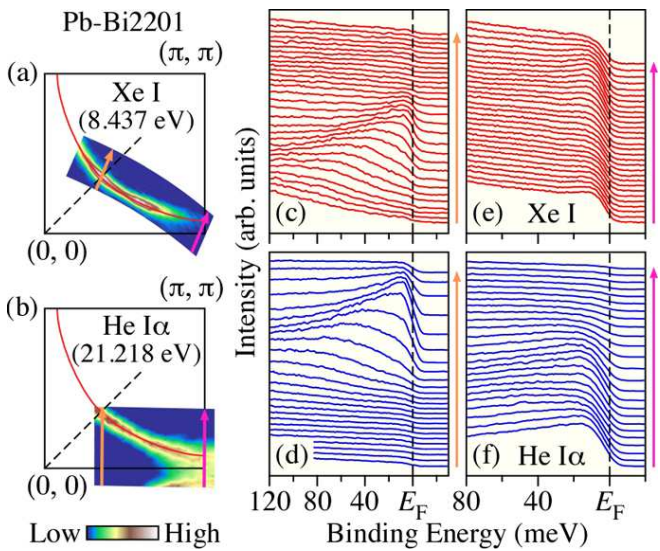}
\end{center}
\caption{
(Color online): (a) and (b) Plot of ARPES intensity at $E_{\rm F}$ for Pb-Bi2201 ($T_{\rm c}$ $\sim$ 21 K) as a function of two-dimensional wave vector measured at 10 K with the Xe-I ($h\nu$ = 8.437 eV) and the He-I$\alpha$ (21.218 eV) lines, respectively.  The intensity at $E_{\rm F}$ was obtained by integrating the spectra within $\pm$15 meV with respect to $E_{\rm F}$.  Red curve represents the Fermi surface determined by smoothly tracing the experimentally determined $k_{\rm F}$ points.  (c) and (d) ARPES spectra measured at 10 K along the orange arrow shown in (a) and (b), respectively.  (e) and (f) Same as (c) and (d), but measured along the pink arrow.
}
\end{figure}

First we present ARPES data in the SC state.  Figures 1(a) and 1(b) show the ARPES intensity plot at $E_{\rm F}$ of Pb-Bi2201 as a function of the two-dimensional wave vector measured with the Xe-I and He-I$\alpha$ lines, respectively.  While the ARPES intensity distribution in the $k$ space is different between these plots likely due to matrix-element effects, we find a nearly identical Fermi-surface shape (red curve) centered at the ($\pi$, $\pi$) point as determined by tracing the Fermi wave vector ($k_{\rm F}$) points.  In the SC state, both the Xe-I and He-I$\alpha$ spectra commonly show a holelike band crossing $E_{\rm F}$ in the nodal region [Figs. 1(c) and 1(d)] and a clear leading-edge shift toward higher binding energy in the antinodal region [Figs. 1(e) and 1(f)].  Although these experimental results suggest the similarity of the basic electronic structure between the He-I$\alpha$ and Xe-I spectra, a closer look further reveals marked differences in the gap behavior.

\begin{figure}[!t]
\begin{center}
\includegraphics[width=3.4in]{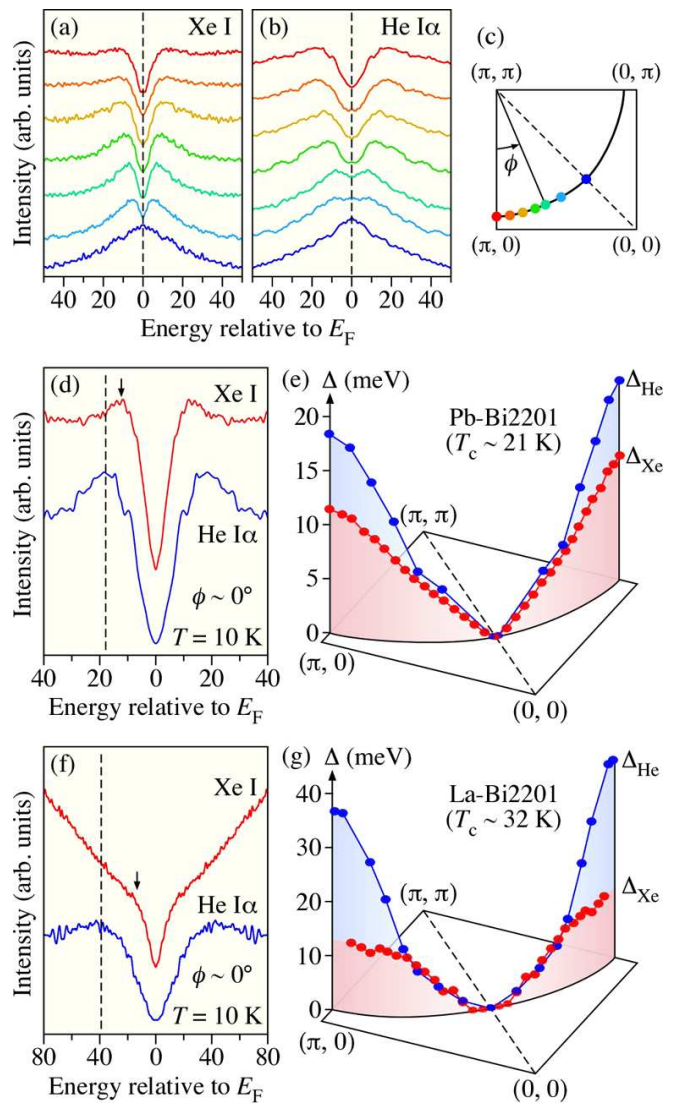}
\end{center}
\caption{
(Color online):  (a) and (b) $k$ dependence of the Pb-Bi2201 ($T_{\rm c}$ $\sim$ 21 K) ARPES spectra at 10 K, measured at various $k_{\rm F}$ points shown by circles in (c), using the Xe-I and He-I$\alpha$ lines, respectively.  The coloring of the spectra is the same as that of the circles in (c).  Each spectrum has been symmetrized with respect to $E_{\rm F}$ to remove the effect of the Fermi-Dirac distribution function.  (c) Schematic Fermi surface and definition of the Fermi-surface angle $\phi$.  (d) Comparison of symmetrized spectra at the antinodal $k_{\rm F}$ point measured with the Xe-I and He-I$\alpha$ lines.  The black arrow and the dashed line denote the peak position for the Xe-I and He-I$\alpha$ spectrum, respectively.  (e) $k$ dependence of the gap size at 10 K obtained with the Xe-I and He-I$\alpha$ lines ($\Delta_{\rm Xe}$ and  $\Delta_{\rm He}$).  The gap size was determined by fitting the symmetrized spectra with the phenomenological gap function convoluted with the energy resolution \cite{Norman}.  (f) and (g) Same as (d) and (e) but measured in La-Bi2201 ($T_{\rm c}$ $\sim$ 32 K).
}
\end{figure}

Figures 2(a) and 2(b) display symmetrized ARPES spectra of Pb-Bi2201 measured at $k_{\rm F}$ points with various Fermi-surface angles $\phi$ at 10 K (below $T_{\rm c}$) with the Xe-I and He-I$\alpha$ lines, respectively.  The gap size, defined by the energy separation between the peak position and $E_{\rm F}$, monotonically increases on going from the nodal (bottom in the panel) to the antinodal (top) regions, which is consistent with the anisotropic gap opening in the SC state.  As also visible in Fig. 2(d), there is a striking difference between the He- and Xe-spectra on the peak position around the antinode, $i.e.$ the peak in the He spectrum is located at much higher binding energy than that of the Xe spectrum (18.5 and 11.5 meV, respectively).  We use $\Delta_{\rm Xe}$ and  $\Delta_{\rm He}$ to note the gap size obtained from the Xe- and He-spectra, respectively.  In Fig. 2(e), we plot estimated  $\Delta_{\rm Xe}$ and  $\Delta_{\rm He}$ at 10 K at various $k_{\rm F}$ points.  The $k$ dependence of  $\Delta_{\rm Xe}$ is well fitted by the  $d_{x^2-y^2}$-wave gap function with a small admixture of higher order component, representing the energy scale of the SC gap \cite{Nakayama}.  Although  $\Delta_{\rm He}$ shows a quantitative agreement with  $\Delta_{\rm Xe}$ near the node, it gradually deviates from  $\Delta_{\rm Xe}$ with approaching the antinode.  Similar trend is also observed in La-Bi2201, whose $T_{\rm c}$ value (32 K) is much higher than Pb-Bi2201 (21 K).  As shown in Fig. 2(f), the difference between  $\Delta_{\rm Xe}$ and  $\Delta_{\rm He}$ (arrow vs dashed line) exceeds 20 meV at the antinode, whereas  $\Delta_{\rm Xe}$ and  $\Delta_{\rm He}$ appear identical near the node.  As visible in Fig. 2(g), the observed significant deviation of  $\Delta_{\rm He}$ from the ideal  $d_{x^2-y^2}$-wave gap function appears similar to previous ARPES results which have been interpreted with two types of energy gaps in {\it different} $k$ regions, $i.e.$ (i) the SC gap which dominates the gap symmetry near the {\it node}, and (ii) the large gap which develops near the {\it antinode} \cite{Tanaka, Kondo1, Terashima}.  The good agreement between  $\Delta_{\rm He}$ and  $\Delta_{\rm Xe}$ near the node in the present ARPES result is consistent with the pairing nature of the gap around the node in the He-spectra.  In addition, the marked difference between  $\Delta_{\rm He}$ and  $\Delta_{\rm Xe}$ near the antinode provides a direct evidence for the presence of two energy gaps below $T_{\rm c}$ (a small gap and a large gap) even in the {\it same} $k$ region, although we cannot completely rule out a possible $k_z$ dependence of the gap size to account for the difference between $\Delta_{\rm He}$ and  $\Delta_{\rm Xe}$.  This observation should be strictly distinguished from previous works reporting the ``two gaps" \cite{Tanaka, Kondo1, Terashima}, in the sense that two gaps appear simultaneously at the antinodal region.

\begin{figure}[!t]
\begin{center}
\includegraphics[width=3.4in]{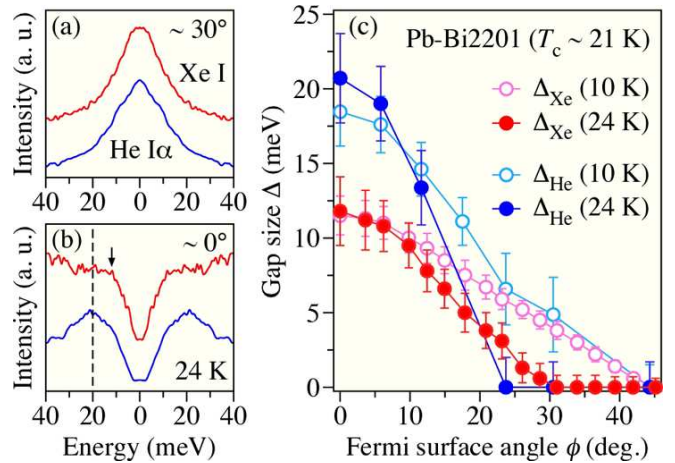}
\end{center}
\caption{
(Color online):  (a) and (b) Photon-energy dependence of symmetrized $k_{\rm F}$ spectra in Pb-Bi2201 ($T_{\rm c}$ $\sim$ 21 K) at 24 K measured at $\phi$ $\sim$30$^{\circ}$ and $\sim$0$^{\circ}$, respectively.  (c) $k$ dependence of the gap size in the SC ($T$ = 10 K) and pseudogap (24 K) states of Pb-Bi2201 measured with the Xe-I and He-I$\alpha$ lines.
}
\end{figure}

To clarify how these gaps evolve into the pseudogap above $T_{\rm c}$, we have performed ARPES measurements at 24 K (just above $T_{\rm c}$) on Pb-Bi2201 with the Xe-I and He-I$\alpha$ lines.  As seen in both sets of data in Fig. 3(a), the symmetrized spectrum near the node shows a single peak at $E_{\rm F}$, while the spectrum at the antinode exhibits spectral weight suppression in the vicinity of $E_{\rm F}$ - a signature of the pseudogap opening [Fig. 3(b)].  In the antinodal region, the characteristic energy scales of the pseudogap are $\sim$12 meV and $\sim$20 meV for the Xe- and He-spectrum, respectively, which are similar to the values of  $\Delta_{\rm Xe}$ and  $\Delta_{\rm He}$ below $T_{\rm c}$ in the antinodal region, as shown in Fig. 3(c).  It is thus inferred that there exist two pseudogaps above $T_{\rm c}$ with precursor-pairing and unknown origin which smoothly evolve from the  $d_{x^2-y^2}$-wave SC gap and the larger gap below $T_{\rm c}$, respectively.  It is emphasized that, although a few previous ARPES results suggested a two-pseudogap-like behavior \cite{He, Kondo2}, the present ARPES result directly demonstrates for the first time the presence of two {\it energy scales} at the antinode.

The present observation solves the contradiction among recent ARPES experiments.  While some studies supported the pairing origin of the pseudogap \cite{Utpal, Feng, Shi, Zhou, Nakayama}, others pointed out that the pseudogap is not directly related to the pairing \cite{Tanaka, Kondo1, Terashima, Ma}.  Such difference is naturally understood by taking into account the presence of two pseudogaps.  Namely, the former studies detected only the small gap and the latter observed mostly the large gap, essentially because of the difference in the experimental conditions such as the photon energy.  In fact, the previously reported pseudogap values of $\sim$15 meV \cite{Feng, Zhou} and $\sim$35 meV \cite{Kondo1, Ma} for La-Bi2201 (which diverse among different groups) agree well with the maximum values of  $\Delta_{\rm Xe}$ and  $\Delta_{\rm He}$, respectively.  In addition, the difference of the gap anisotropy in the pseudogap phase of La$_{1.875}$Ba$_{0.125}$CuO$_4$ \cite{He, Valla} can also be explained within the two-pseudogap picture.

Finally, we discuss the implication of the observed photon-energy dependence.  We revealed that the measurements using the Xe-I ($h\nu$ = 8.437 eV) and the He-I$\alpha$ (21.218 eV) lines are sensitive to the small gap and the large gap, respectively.  One explanation of such behavior is that the two different gaps suffer different matrix-element effects during the photo-excitation process and they can be selectively observed by specific conditions of the photon energy.  This explanation may be valid if there are two different bands producing the small and the large gaps, respectively, as the bilayer-split bands in Bi2212 obey different matrix elements \cite{Borisenko}.  On the other hand, Bi2201 is a single-layered system and there would be a single band near $E_{\rm F}$.  In this case, the appearance of two energy scales on the single coherent quasiparticle band may be explained by the idea that the large gap is not a complete gap but rather a soft gap \cite{Ma}, and the remaining density of states within the large gap contributes to the formation of the small gap.  It is also possible to attribute the large and the small gaps to the incoherent and the coherent parts of the spectral function.  In either case, the two gaps basically arise from a single-band spectral function and their intensity ratio would not depend on the photon energy.  Hence we think that the present observation may not be simply explained by the matrix-element effect.  Another explanation is that the difference between the He- and Xe-spectrum originates from the surface / bulk sensitivity.  In this case, it is inferred that the large gap, which seems not directly related to the superconductivity, is either (i) an extrinsic feature stabilized at the surface or (ii) an intrinsic feature in bulk with much pronounced influence at the surface.  On the other hand, the small gap, which is closely related to the pairing, would reflect bulk properties because electrons excited with the Xe-I line have a relatively long escape depth (20-40 \AA) as compared to that excited with the He-I$\alpha$ line (5-10 \AA) \cite{Souma}.  The bulk nature of the small gap is also supported by a basic agreement between  $\Delta_{\rm Xe}$ in La-Bi2201 ($\sim$14 meV) and an energy scale observed in the $B_{1g}$ Raman spectrum ($\sim$17 meV) \cite{Sugai}.  While most of previous results on Bi2201 \cite{Kondo1, Ma, Zhou, Nakayama, Kondo2} agree with the expectation that the spectral weight related to the small gap feature is enhanced as the photon energy is lowered ($i.e.$ the photoelectron escape depth becomes longer), there is one exceptional result which shows the small gap feature at the antinode even with $h\nu$ = 22.5 eV in optimally doped La-Bi2201 with a zero residual resistivity \cite{Feng}.  Since the authors reported that the small gap disappears in another optimally doped sample with a finite residual resistivity \cite{Feng}, the disorder effect may be an essential ingredient in suppressing the small gap component and also in causing the difference in the electronic states between surface and bulk.  In any cases, the pairing interaction is essential in realizing the origin of the pseudogap, and we conclude that the scenario assuming the opening of a single competing pseudogap is insufficient for the correct understanding of the pseudogap phenomena in cuprates.

\section{IV. CONCLUSION}
In conclusion, we have performed high-resolution ARPES study of Bi2201 by using the Xe and He discharge lamps.  The result clearly shows the presence of two energy scales in the antinodal region below and above $T_{\rm c}$, indicating the existence of two different pseudogaps.  We have concluded that the smaller pseudogap originates from the precursor pairing above $T_{\rm c}$, while the larger pseudogap is not directly related to the superconductivity.  The present findings put a strong constraint in modeling the pseudogap phenomena of cuprates.

\section{ACKNOWLEDGMENTS}
This work was supported by grants from JST-CREST, JSPS, MEXT of Japan, NSF of US, and MOST of China..

\end{document}